\documentclass[
    ,final            
  ]
  {aipproc}

\layoutstyle{6x9}

\begin{document}

\title{Evolved solar systems in Praesepe}


\author{S. L. Casewell}{
  address={Dept Physics and Astronomy, University of Leicester, Leicester, LE1 7RH, UK}
}

\author{M. R. Burleigh}{
  address={Dept Physics and Astronomy, University of Leicester, Leicester, LE1 7RH, UK}
}

\author{P. D. Dobbie}{
  address={AAO, PO Box 296, Epping, NSW 1710, Australia}
  }
\author{R. Napiwotzki}{
  address={Centre for Astrophsyics Research, University of Hertfordshire, Hatfield, AL10 9AB, UK}
}

\begin{abstract}
We have obtained near-IR photometry for the 11 Praesepe white dwarfs, to search for an excess indicative of a dusty debris disk.
All the white dwarfs are in the DAZ temperature regime, however we find no indications of a disk around any white dwarf. We have, however determined
that the radial velocity variable white dwarf WD0837+185 could have an unresolved  T8 dwarf companion that would not be seen as a near-IR excess. 
\end{abstract}

\maketitle

\section{Introduction}


There are $\approx10$ dusty debris disks known to exist around white dwarfs.  The disks are found around relatively cool white dwarfs ($9,000<T<22,000$K) whose atmospheres are polluted with heavy elements (the so-called
DAZ white dwarfs). The disks provide the obvious reservoir for the white dwarf to accrete this material, which otherwise would sink from the
atmosphere on a timescale of days.  DAZs are identified
through high resolution, high S/N optical spectra in which lines of Ca, Si and Fe can be detected.

\cite{zuckerman03} estimated that 20-25 per cent of single  DA white dwarfs show Ca \textsc{ii} K lines, indicating they are DAZ. This statistic has been put into question, 
however by \cite{koester05} who studied 478 DA white dwarfs with 10 000 K $\geq$ T$_{\rm eff} \geq$30 000 K and found 24 DAZs, 6 of which had been discovered by 
\cite{zuckerman03}. This put the fraction of DA white dwarfs that are DAZ at 0.5 per cent. It was suggested by \cite{koester05} that this discrepancy has occurred as
the \cite{zuckerman03} sample was mainly of objects with T$_{\rm eff}$$<$ 10 000 K, where the Ca absorption lines at lower abundances are easiest to detect, so their sample was biased towards the DAZs.

\cite{kilic08} studied a sample of 37 DAZ white dwarfs using IRTF and $Spitzer$, of which 7 had dusty debris disks. They tentatively estimate the fraction of DAZ that harbour disks at $\approx$ 20 per cent, although this is obviously from a small sample.

Open star clusters are ideal places within which to search for white dwarfs with disks as all cluster members have a known age. Therefore we can calculate the cooling age of any white dwarf, and hence the mass of the progenitor star. We recently investigated the 
white dwarf members of the moderately rich nearby Praesepe open cluster, measuring their effective temperatures and gravities \cite{casewell09, dobbie06}. 
We identified that WD0837+218 has a radial velocity that is inconsistent with it being a cluster member, but have included it in the sample here for completeness.
All eleven white dwarfs have temperatures between 14 5000 K and 22 000 K with log g between 8.1 and 8.45, in the DAZ range, although the values for the magnetic white dwarf WD0836+201 should be regarded with some scepticism as the temperature fitting is unreliable. The above statistics suggest we should find a 0-2 DAZs in our sample.

At a distance of 177$^{+10.3}_{-9.2}$ pc (as determined from Hipparcos 
measurements, \cite{mermilliod97}), Praesepe  is one of the closest star clusters. It is slightly metal rich with respect to the Sun ([Fe/H] = +0.11, \cite{an07}).  Indeed, as both the metallicity and the kinematics of Praesepe are similar to those of the Hyades, the former is often touted as a 
member of the Hyades moving group and therefore is assumed have an age comparable to the latter, $\tau$=625$\pm$50 Myr (e.g. \cite{claver01}). We note that this age for the Hyades was derived by comparing model isochrones generated from slightly metal enhanced (Z=0.024) stellar models which included moderate convective overshooting
to the colors and magnitudes of a sample of cluster members selected using Hipparcos astrometric data \cite{perryman98}. 

\section{Acquisition and reduction of data}

The data were acquired between 6/12/08 and 24/01/09 using the United Kingdom
Infrared Telescope (UKIRT) and the UKIRT Fast Track Imager (UFTI).

The data were taken with total exposure times of 3000 s in the $K$
band, 1200 s  in the $H$ band and 600 s in the $J$ band. All images were taken
using a 5 point jitter pattern with 60 s exposures. A standard star from the UKIRT faint standards (Casali \& Hawarden, 1992: UKIRT Newsletter, 4, 33)  was also
observed for each observation with a total time of 15 s in the $H$ and $K$
band (5 3 s exposures) and 40 s in the $J$ band (5 8 s exposures).

The data were reduced using the \textsc{starlink} based \textsc{orac-dr} pipeline with the recipe \textsc{jitter\_self\_flat}
 which performs the dark correction and creates and applies a sky flat field to the images before mosaicing them.
The aperture for the standard star was set to 5 times the FWHM, and an aperture correction was derived using bright stars in the target field. This was performed using the \textsc{iraf digiphot} package \textsc{qphot}.
The photometry can be seen in Table \ref{gcs}.
\begin{table}
\caption{\label{gcs}The infrared  $J$, $H$ and $K$ photometry for the Praesepe white dwarfs.The UKIDSS photometry is give in italics.}
\begin{tabular}{lccccc}
\hline
Name & $J$&$H$&$K$\\
\hline
WD0833+194&18.623$\pm$0.039 &18.746$\pm$0.063&18.553$\pm$0.017\\

WD0836+197&- &- &18.283$\pm$0.022\\
WD0836+199&$\mathit{18.602\pm0.066}$  &18.685$\pm$0.025&18.892$\pm$0.038\\

WD0836+201&$\mathit{18.195\pm0.0455}$ &$\mathit{18.098 \pm0.106}$&18.476$\pm$0.03\\
WD0837+185& 18.609$\pm$0.028&18.568$\pm$0.019&18.605$\pm$0.019\\
&18.689$\pm$0.016&18.860$\pm$0.06&18.70$\pm$0.02\\
&18.765$\pm$0.03&-&18.69$\pm$0.028\\
&-&-&18.638$\pm$0.033\\
WD0837+199&-&$\mathit{17.305 \pm0.05}$&   $\mathit{16.481\pm0.047}$\\
  
WD0837+218&$\mathit{18.333 \pm0.0833}$& $\mathit{18.434\pm0.116}$&$\mathit{18.448\pm0.224}$\\

WD0840+190&18.516$\pm$0.027 &$\mathit{18.752\pm0.177}$&18.808$\pm$0.026\\
 
WD0840+200& 18.574$\pm$0.025&$\mathit{18.390 \pm0.1225}$&18.525$\pm$0.02\\
WD0840+205 &$\mathit{18.616 \pm0.065}$& $\mathit{18.549\pm0.137}$&   -\\
WD0843+184 &18.747$\pm$0.029&$\mathit{18.595\pm0.135}$&18.786$\pm$0.029\\
\hline
\end{tabular}
\end{table}
\section{Results}

Using T$_{\rm eff}$ and log g from \cite{casewell09}, we generated WD models
using \textsc{tlusty} and \textsc{synspec} that extend
from 0.3 to 2.5 microns, covering the Sloan Digital Sky Survey $u$, $g$, $r$, $i$, $z$ and the near-IR  photometry range.

All the white  dwarfs in this sample have SDSS photometry except for WD0836+199
which is too close to a nearby bright star as discussed in \cite{casewell09}. The SDSS photometry $g$ band has been estimated for this star using its
T$_{\rm eff}$ and log g.
For each object, the  model was normalized to the $i$ band as using $g$ and $r$ led to underpredictions of the $JHK$ photometry due to several deep absorption lines in the $g$ band, and H$\alpha$ in the $r$ band.

Some of these white dwarfs also have $Z$, $Y$, $J$, $H$ and $K$ from the UKIRT Infrared Sky Survey (UKIDSS) Galactic Cluster Survey (GCS) DR6 and these have also been included where relevant.

None of the Praesepe white dwarfs possesses a detectable debris disk except from WD0837+199, which shows a clear near-IR excess in the UKIDSS $H$ and $K$ bands (Figure \ref{wd0837}, left panel) and also in Spitzer IRAC photometry at 4.5 and 8 microns \cite{gaspar09}. However, our deeper UFTI images clearly show a nearby, red galaxy  (Figure \ref{wd0837}, right panel), implying this excess is unlikely to be due to the white dwarf.


\begin{figure}[ht]
\begin{minipage}[b]{0.25\linewidth}
\centering
\includegraphics[height=.3\textheight, angle=270]{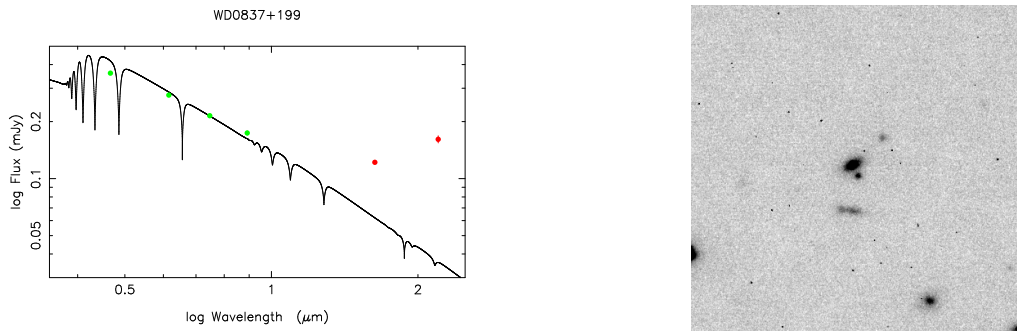}
\caption{Left:$g$, $r$, $i$, $z$, $H$, $K$ magnitudes for WD0837+199 plus a model spectrum of a 14,000K DA white dwarf. An excess can clearly be seen.
Right:$K$ band image of WD0837+199. North is up and East  is left. The red galaxy can be seen to the north of the white dwarf in the centre of the image.}
\label{wd0837}
\end{minipage}
\hspace{5.0cm}
\begin{minipage}[b]{0.25\linewidth}
\centering
\includegraphics[height=.2\textheight,  angle=270]{slcasewell2}
\label{wd0837}
\end{minipage}
\end{figure}
\begin{figure}
  \includegraphics[height=.4\textheight, angle=270]{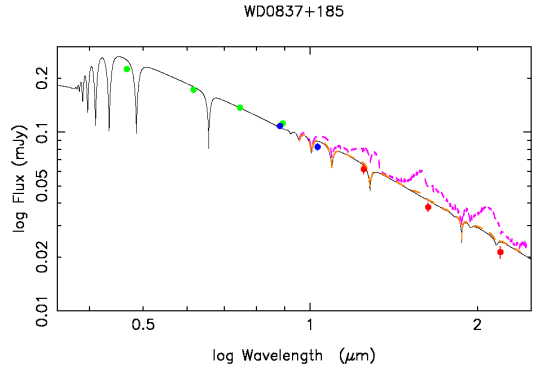}
  \caption{{\label{wd+bd}} The combined model spectrum of a 14,000K DA WD+T5 brown dwarf at the distance of Praesepe has been plotted in purple (dotted line), and a WD+T8 in orange (dashed line) as well as SDSS $g$, $r$, $i$, $z$, UKIDSS $Z$, $Y$ and UFTI $J$, $H$ and $K$ magnitudes for WD0837+185.}
\end{figure}

One other object of note is WD0837+185, a radial velocity variable. Our near-infrared photometry shows no excess although Figure \ref{wd+bd} shows that an unresolved T8 (M$\approx$25M$_{\rm Jup}$ from the RV curve) brown dwarf companion could be hidden by the white dwarf. We require further data before we can make any firm conclusions however.

\section{Conclusions}
\cite{kilic08} suggest that up to 25 per cent of externally polluted DAZ white dwarfs show an infrared excess indicative of a dust disk, and in the absence of prior knowledge of metallicity, \cite{farihi09} expect 1-3 per cent of all single white dwarfs with cooling ages less than $\approx$0.5 Gyr to possess dust disks. Hence, it is not necessarily surprising that we have not detected a dust disk in Praesepe. However, we are able to place constraints on the presence of substellar companions. These white dwarfs have evolved from late B-type main sequence stars (2.9-3.5 M$_{\odot}$ \cite{casewell09}), spectral types that are not observed in radial velocity searches for substellar and planetary companions and we have determined that a T8 brown dwarf cannot be detected in the UKIDSS and UFTI observations of WD0837+185. Such a brown dwarf has a mass of 25 M$_{\rm Jup}$ at 625 Myr. Hence, these limits can be combined with the results of radial velocity searches for substellar companions to place limits on the formation of brown dwarfs as binary companions to late B-stars. 

\begin{theacknowledgments}
The United Kingdom Infrared Telescope is operated by the Joint Astronomy Centre on behalf of the Science and Technology Facilities Council of the U.K.
\end{theacknowledgments}

\bibliographystyle{aipproc}   

\bibliography{slcasewell}

\hyphenation{Post-Script Sprin-ger}
\begin{thebibliography}{11}
\expandafter\ifx\csname natexlab\endcsname\relax\def\natexlab#1{#1}\fi
\providecommand{\enquote}[1]{``#1''}
\expandafter\ifx\csname url\endcsname\relax
  \def\url#1{\texttt{#1}}\fi
\expandafter\ifx\csname urlprefix\endcsname\relax\def\urlprefix{URL }\fi
\providecommand{\eprint}[2][]{\url{#2}}

\bibitem[{Zuckerman} et~al.(2003)]{zuckerman03}
B.~{Zuckerman}, et~al., \emph{\apj} \textbf{596}, 477 (2003).

\bibitem[{Koester} et~al.(2005)]{koester05}
D.~{Koester}, et~al., \emph{\aap} \textbf{432}, 1025 (2005).

\bibitem[{Kilic} et~al.(2008)]{kilic08}
M.~{Kilic}, et~al., \emph{\aj} \textbf{136}, 111 (2008).

\bibitem[{Casewell} et~al.(2009)]{casewell09}
S.~L. {Casewell}, et~al., \emph{\mnras} \textbf{395}, 1795 (2009).

\bibitem[{Dobbie} et~al.(2006)]{dobbie06}
P.~D. {Dobbie}, et~al., \emph{\mnras} \textbf{369}, 383 (2006).

\bibitem[{Mermilliod} et~al.(1997)]{mermilliod97}
J.~{Mermilliod}, et~al., \enquote{{The Distance of the Pleiades and Nearby
  Clusters},} in \emph{Hipparcos - Venice '97}, edited by {R.~M.~Bonnet,
  E.~H{\o}g, P.~L.~Bernacca, L.~Emiliani, A.~Blaauw, C.~Turon, J.~Kovalevsky,
  L.~Lindegren, H.~Hassan, M.~Bouffard, B.~Strim, D.~Heger, M.~A.~C.~Perryman,
  \& L.~Woltjer}, 1997, vol. 402 of \emph{ESA Special Publication}, p. 643.

\bibitem[{An} et~al.(2007)]{an07}
D.~{An}, et~al., \emph{\apj} \textbf{655}, 233 (2007).

\bibitem[{Claver} et~al.(2001)]{claver01}
C.~F. {Claver}, et~al., \emph{\apj} \textbf{563}, 987 (2001).

\bibitem[{Perryman} et~al.(1998)]{perryman98}
M.~A.~C. {Perryman}, et~al., \emph{\aap} \textbf{331}, 81 (1998).

\bibitem[{G{\'a}sp{\'a}r} et~al.(2009)]{gaspar09}
A.~{G{\'a}sp{\'a}r}, et~al., \emph{\apj} \textbf{697}, 1578 (2009).

\bibitem[{Farihi} et~al.(2009)]{farihi09}
J.~{Farihi}, M.~{Jura}, and B.~{Zuckerman}, \emph{\apj} \textbf{694}, 805
  (2009).

\end{thebibliography}


\end{document}